\documentclass[twocolumn,prl,showpacs]{revtex4}
\usepackage{amsmath}
\usepackage{graphicx}
\newcommand{\ket}[1]{\ensuremath{|#1\rangle}}

\newcommand{\half}{\mbox{$\frac{1}{2}$}}

\newcommand{\NOT}{\textsc{not}}

\newcommand{\CNOT}{controlled-\NOT}

\begin{document}
\title{Robust Ising Gates for Practical Quantum Computation}
\author{Jonathan A. Jones}\email{jonathan.jones@qubit.org}
\affiliation{Centre for Quantum Computation, Clarendon Laboratory,
University of Oxford, Parks Road, OX1 3PU, United Kingdom}
\date{\today}
\begin{abstract}
I describe the use of techniques based on composite rotations to
combat systematic errors in controlled phase gates, which form the
basis of two qubit quantum logic gates. Although developed and
described within the context of Nuclear Magnetic Resonanace (NMR)
quantum computing these sequences should be applicable to any
implementation of quantum computation based on Ising couplings.
In combination with existing single qubit gates this provides a
universal set of robust quantum logic gates.
\end{abstract}
\maketitle

Quantum computers \cite{bennett00} use quantum mechanical effects
to implement algorithms which are not accessible to classical
computers, and thus are able to tackle otherwise intractable
problems \cite{shor99}. They are extremely vulnerable to the
effects of errors, and considerable effort has been expended on
correcting random errors arising from decoherence processes
\cite{shor95, steane96, steane99}.  It is, however, also important
to consider the effects of systematic errors, which arise from
reproducible imperfections in the experimental apparatus used to
implement quantum computations.

It makes sense to consider systematic errors as some of them can
be tackled relatively easily.  In the context of Nuclear Magnetic
Resonance (NMR) quantum computation \cite{cory96, cory97,
gershenfeld97, jones98c, jones01a, vandersypen01} the concept of
composite rotations (also called composite pulses) \cite{ernst87,
freeman97b, levitt86, wimperis94} has been used to tackle both
off-resonance effects \cite{cummins00} and more recently pulse
length errors \cite{cummins02}.  All these results, however, have
been concerned with single qubit gates, and while these play a
central role in quantum logic circuits it is also vital to
consider two qubit controlled logic gates, such as the \CNOT\
gate.

The combination of a complete set of single qubit gates and the
\CNOT\ gate is \emph{universal} for quantum computation
\cite{barenco95}, that is any quantum logic circuit can be built
entirely from single qubit gates and \CNOT\ gates.  There are,
however, other two qubit gates which can be used to form a
universal set, most notably the controlled phase-shift
(controlled-$\sigma_z$) gate \cite{jones01a}.  This performs the
transformation
\begin{equation}
\ket{1}\ket{1}\longrightarrow -\ket{1}\ket{1}
\end{equation}
while leaving other eigenstates unchanged, and so is described by
the matrix
\begin{equation}
\begin{pmatrix}1&0&0&0\\0&1&0&0\\0&0&1&0\\0&0&0&-1\end{pmatrix}.
\label{eq:cps}
\end{equation}
Applying Hadamard gates to the target spin before and after the
controlled-$\sigma_z$ gate converts it to a controlled-$\sigma_x$
gate, that is \CNOT\ \cite{jones01a}.

The controlled phase-shift gate can itself be considered as a
derivative of the Ising ($zz$) coupling gate
\begin{equation}
\begin{pmatrix}e^{i\phi/2}&0&0&0\\0&e^{-i\phi/2}&0&0\\0&0&e^{-i\phi/2}&0\\0&0&0&e^{i\phi/2}\end{pmatrix}
\label{eq:zz}
\end{equation}
which corresponds to evolution under the Ising
($\sigma_{1z}\sigma_{2z}$) Hamiltonian; Eqns. \ref{eq:cps} and
\ref{eq:zz} are related by single qubit $z$ rotations and a global
phase shift \cite{jones01a}.  The $z$ rotations can be adsorbed
into abstract reference frames \cite{knill00} and the global phase
shift has no physical significance, and thus the two equations are
essentially equivalent.

The Ising coupling is of great experimental importance, as it
provides the basic quantum logic gate for many proposed
implementations of quantum computing \cite{cory96, lloyd93,
ioffe99, cirac00, briegel01,raussendorf01}. Its role is
particularly clear in NMR systems, where the scalar coupling ($J$
coupling) takes the Ising form in the weak coupling limit
\cite{ernst87}, and from this point I will adopt the product
operator notation widely used in NMR \cite{ernst87, jones01a,
sorensen83}.  However the underlying principles remain applicable
in any system based on Ising couplings.

Consider a system of two spin-\half\ nuclei, $I$ and $S$.  The
Ising gate (Eq.~\ref{eq:zz}) is implemented by free evolution
under the $J$ coupling Hamiltonian
\begin{equation}
\mathcal{H}_{IS}=\pi J\, 2I_zS_z
\end{equation}
for a time $\tau=\phi/\pi J$, where $J$ is the coupling strength
and $\phi$ is the desired evolution angle. In order to implement
accurate controlled phase-shift gates by this means it is clearly
necessary to know $J$ with corresponding accuracy.

Obtaining accurate values of $J$ for individual spin--spin
couplings is easy in small molecules, but is much more difficult
in larger systems.  In particular, consider an array of qubits
coupled by Ising interactions, with $J$ couplings that are
nominally identical but in fact differ from one another as a
result of imperfections in the lattice.  In such a system it is
desirable to be able to perform some accurately known $zz$
evolution over a \emph{range} of values of $J$.  While this might
appear tricky, it is in fact easy to achieve using composite pulse
techniques.

The problem of performing accurate $zz$ rotations is conceptually
similar to that of correcting for pulse length errors in single
qubit gates \cite{cummins02}, and the solutions are closely
related.  I begin by describing existing techniques for tackling
pulse length errors.  While a wide variety of composite pulse
sequences have been developed within the context of NMR
experiments \cite{levitt86}, most of these are not suitable for
use in quantum computers as they make assumptions about the form
of the initial state.  Instead it is necessary to use Class A
composite pulses \cite{levitt86} which work well for any initial
state.  Two suitable families of composite pulses have been
described: the \textsc{scrofulous} pulse sequences of Cummins
\emph{et al.} \cite{cummins02}, and the BB1 family, initially
described by Wimperis \cite{wimperis94}, and subsequently
developed by Cummins \emph{et al.} \cite{cummins02}.  As the
performance of the BB1 sequence is much better than that of
\textsc{scrofulous} I will concentrate on the time symmetrized
variant of BB1.

Consider a $\theta_0$ pulse, which implements a single qubit
rotation around the $x$ axis on the Bloch sphere \cite{ernst87}.
In an ideal system the propagator will be
\begin{equation}
U=\exp[-i\theta I_x]
\end{equation}
but in the presence of a fractional error $\epsilon$ in the
strength of the driving field the actual propagator will be
\begin{equation}
V=\exp[-i\theta(1+\epsilon)I_x]. \label{eq:naivex}
\end{equation}
The accuracy of this experimental implementation can be assessed
using the propagator fidelity
\begin{equation}
\mathcal{F}=\frac{|\text{Tr}(VU^\dag)|}{\text{Tr}(UU^\dag)}
\end{equation}
(note that it is necessary to take the absolute value of the
numerator as $U$ and $V$ could in principle differ by an
irrelevant global phase shift).  In this case
\begin{equation}
\mathcal{F}=\cos\left(\frac{\epsilon\theta}{2}\right)\approx
1-\frac{\epsilon^2\theta^2}{8}.
\end{equation}
A better approach is to replace the naive implementation,
Eq.~\ref{eq:naivex}, with the composite pulse sequence
\begin{equation}
(\theta/2)_0\,\pi_{\phi_1}\,2\pi_{\phi_2}\,\pi_{\phi_1}\,(\theta/2)_0
\label{eq:BB1}
\end{equation}
with $\phi_2=3\phi_1$ and $\phi_1=\arccos(-\theta/4\pi)$.  This
gives a fidelity expression
\begin{equation}
\mathcal{F}\approx1-\frac{\epsilon^6(\theta^6-14\pi^2\theta^4-32\pi^4\theta^2)}{9216}
\end{equation}
in which the second \emph{and} fourth order error terms have been
completely removed \cite{cummins02}.

\begin{figure}
\includegraphics{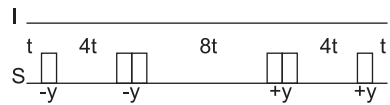}
\caption{Pulse sequence for a robust Ising gate to implement a \CNOT\ gate.
Boxes correspond to single qubit rotations with rotation angles of
$\phi=\arccos(-1/8)\approx97.2^\circ$ applied along the $\pm y$ axes as
indicated; time periods correspond to free evolution under the Ising coupling,
$\pi J\,2I_zS_z$ for multiples of the time $t=1/4J$.  The naive Ising gate
corresponds to free evolution for a time $2t$.} \label{fig:pulses}
\end{figure}

A very similar approach can be used to tackle systematic errors in
Ising coupling gates; in effect Ising coupling corresponds to
rotation about the $2I_zS_z$ axis, and errors in $J$ correspond to
errors in a rotation angle about this axis.  These can be
parameterised by the fractional error in $J$:
\begin{equation}
\epsilon=\frac{J_{\textit{real}}}{J_{\textit{nominal}}}-1.
\end{equation}
Such errors can be overcome by rotating about a sequence of axes
tilted from $2I_zS_z$ towards another axis, such as $2I_zS_x$.
Defining
\begin{equation}
\theta_\phi\equiv\exp[-i\times\theta\times(2I_zS_z\cos\phi+2I_zS_x\sin\phi)]
\end{equation}
allows the naive sequence $\theta_0$ to be replaced by
Eq.~\ref{eq:BB1} as before. The tilted evolutions can be achieved
by sandwiching a $2I_zS_z$ rotation (that is, free evolution under
the Ising Hamiltonian) between $\phi_{\mp y}$ pulses applied to
spin $S$ \cite{ernst87}.  For the case that $\theta=\pi/2$ (which
forms the basis of the \CNOT\ gate) the final sequence takes the
form shown in Fig.~\ref{fig:pulses}.

Clearly it is vital that any robust implementation of a quantum
gate must be built from components that can themselves be
implemented robustly.  The robust Ising gate uses only two
components: single qubit rotations around the $\pm y$ axes, for
which robust versions are already known \cite{cummins02}, and
periods of evolution under the Ising coupling.  Note that the five
time periods must have lengths in the integer ratios $1:4:8:4:1$,
but it is not necessary to accurately control the absolute length
of the periods, as errors in absolute lengths are equivalent to
errors in the value of $J$.

\begin{figure*}
\raisebox{135pt}{(a)}\quad\includegraphics*{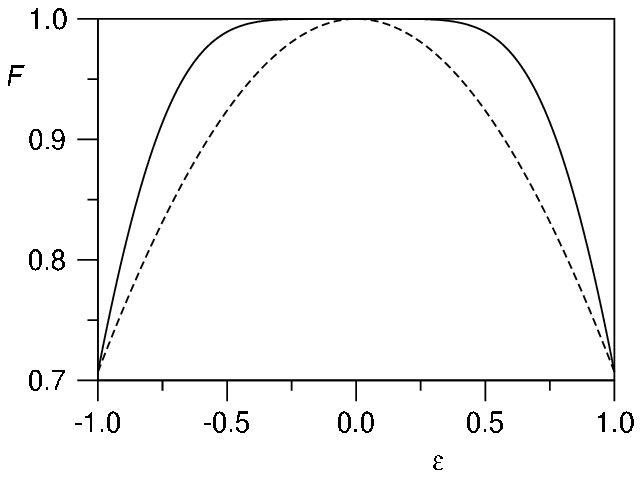} \qquad
\raisebox{135pt}{(b)}\quad\includegraphics*{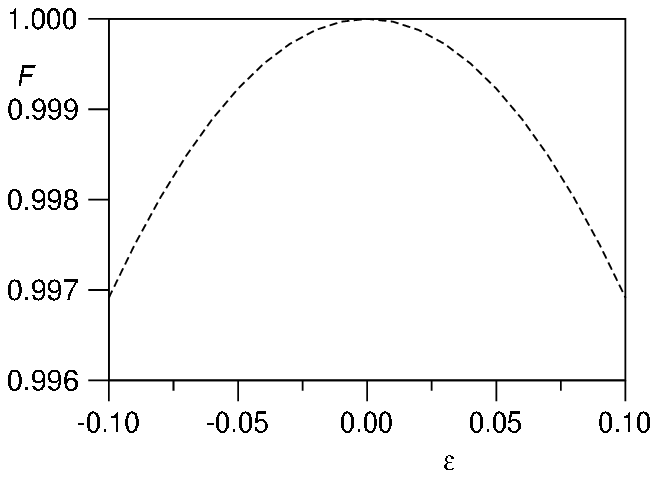}
\caption{Fidelity of naive (dashed line) and robust BB1 (solid
line) Ising gates as a function of $\epsilon$, the fractional
error in $J$. Plots are shown for (a) errors in the range of $\pm
100\%$ and (b) errors in the range of $\pm 10\%$.  Note that in
plot (b) the BB1 line is indistinguishable from perfection.}
\label{fig:fidelity}
\end{figure*}

The fidelity gain achieved for coupling gates by this approach is
\emph{identical} to that achieved for single qubit rotations.  In
particular for the case that $\theta=\pi/2$ the naive fidelity is
\begin{equation}
\mathcal{F}\approx 1-\frac{\epsilon^2\pi^2}{32}
\end{equation}
while the BB1 approach gives
\begin{equation}
\mathcal{F}\approx 1-\frac{\epsilon^663\pi^6}{65536}.
\end{equation}
Fidelity plots are shown in Fig.~\ref{fig:fidelity}.
Clearly the BB1 approach compensates extremely well for small
errors in $J$ values, especially within the range $\pm10\%$.  Over
this range the infidelity of the BB1 sequence is always less that
one part in $10^6$.  To achieve comparable fidelity with a naive
gate it is necessary to determine the value of $J$ to better than
$0.2\%$.  Thus the robust gate can achieve an infidelity of
$10^{-6}$ over a range of $J$ values more than 50 times wider than
the naive gate. If even higher fidelities are desired the
improvement provided by the robust gate is even greater.  In
combination with existing robust single qubit gates, the robust
Ising gate provides a complete set of robust gates for quantum
computation within the Ising model.

\begin{acknowledgments}
I thank the Royal Society of London for financial support.  I am
grateful to S.~Benjamin, L.~Hardy, M.~Mosca and A.~M. Steane for
helpful discussions.
\end{acknowledgments}

\end{document}